\documentclass[referee]{aastex}

\shorttitle{Self-Consistent Disk Response to Elliptical Halo Potential}
\shortauthors{Jog}
\begin{document}

\title{ SELF-CONSISTENT RESPONSE OF A GALACTIC DISK TO AN ELLIPTICAL 
  PERTURBATION HALO POTENTIAL}  

\author{Chanda J. Jog}

\affil{Department of Physics, Indian Institute of
   Science, Bangalore 560012, India, \\email: cjjog@physics.iisc.ernet.in}

\begin{abstract}
We calculate the self-consistent response of an axisymmetric galactic disk 
perturbed by an elliptical halo potential of harmonic number
$m=2$, and obtain the net disk ellipticity. Such a potential is commonly 
expected to arise due to galactic tidal encounter and also during the galaxy 
formation process.  
The self-gravitational potential corresponding to the  self-consistent, 
non-axisymmetric density response of the disk is obtained by inversion of 
Poisson equation for a thin disk. This response potential is shown 
to oppose the perturbation potential, because physically
the disk self-gravity resists the imposed potential. This
results in a reduction in the net ellipticity of the perturbation halo
potential in the disk plane. The reduction factor denoting this decrease is 
independent of the strength of the perturbation potential, and has a
typical minimum value of  $\sim  0.75 - 0.9 $   for a wide range of galaxy 
parameters. The reduction is most important at 1.4 exponential
disk scale lengths and is progressively
less so at higher radii. For the solar neighborhood region of the 
Galaxy,  the reduction factor is 0.8 .
  Beyond twice the Holmberg radius, the 
reduction is negligible, and there the disk asymmetry in the atomic hydrogen 
gas traces the true ellipticity of the halo potential.
The reduction is negligible at all
radii for higher harmonics ($m \geq 3$) of the halo potential.
 
On correcting for the negative disk response, the {\it true} ellipticity of 
the halo potential for a typical spiral galaxy is shown to be higher by 
$\sim 20 \% $ than the typical halo ellipticity of $\leq 0.1$ deduced in the 
literature from observations of isophotal or kinematical asymmetry of disks.
\end{abstract}


\section{INTRODUCTION}

It is now realized that the disks of spiral galaxies display a
rich variety of non-axisymmetry in their light and hence mass distribution.
Lopsided galaxies such as M 101 and NGC 1637 show a global asymmetry, with a 
much larger spatial extent on one side than on the other, as
seen optically (Sandage 1961), and particularly strongly in the
atomic hydrogen gas (Baldwin, Lynden-Bell, \& Sancisi 1980).  
Thus the disk mass distribution in these is characterized by $m = 1$ where $m$ 
is the harmonic number or the azimuthal number of the Fourier
component being studied. For the elliptical case ($m = 2 $),
similar globally asymmetric distributions with a constant phase with radius  
are not easy to discern. This is because it is difficult to separate the 
projection effects from the intrinsic ellipticity of the disk, and hence this
needs a careful photometric study of nearly face-on galaxies as done for 
example by Rix \&
Zaritsky (1995). On the other hand the $m = 2$ spiral features with a
phase varying with radius say as in M 81 or M 51 (Sandage 1961)
are well studied, and indeed are responsible for the `spiral' nomenclature
for these galaxies. These earlier studies were done in the
blue-band, and hence the features show a much stronger contrast than the
asymmetry in the underlying stellar mass distribution which is
harder to measure.

Recent near-IR observations allow one to measure the mass asymmetry in the 
underlying, old stellar disk population (Block et al. 1994, Rix \& 
Zaritsky 1995, Zaritsky \& Rix 1997, Rudnick \& Rix 1998, Kornreich, Haynes \&
Lovelace 1998). Rix \&  Zaritsky (1995) have given a quantitative measure of 
the amplitudes of the components with various harmonic numbers and their 
radial variations. The amplitudes of the higher harmonics ($m > 3$) are 
observed to be generally smaller than those for $m = 1,2$ or $3$. The 
asymmetry at a higher radius
beyond the optical disk or the Holmberg radius is better studied using the HI
as a tracer as shown for the $m=1$ case, with mapping (Baldwin
et al. 1980) and with the global velocity profiles (Richter \& Sancisi 1994; 
Haynes et al. 1998). The disk asymmetry can also be deduced from
the kinematic study of the HI velocity field observations
(Schoenmakers, Franx, \& de Zeeuw 1997).

Despite the small value of the observed disk asymmetry of a few \% ,
it is becoming an interesting topic for study since the disk asymmetry
is tied into the asymmetry of the halo and indeed provides a quantitative
measure or diagnostic of the halo asymmetry. 
This is because the global asymmetry in the disk is attributed as a response to
the halo distortion ( e.g., Weinberg 1995, Jog 1997).
Thus the halo asymmetry can be deduced from the observed isophotal 
(Rix \& Zaritsky 1995), or the kinematic (Franx \& de Zeeuw 1992) disk 
asymmetry. Binney (1978) first proposed that a galaxy halo 
would be  non-axisymmetric / triaxial, and studied 
its effect on the embedded disk. 
The halo asymmetry could arise either due to tidal interactions 
between galaxies  (Wienberg 1995), or the triaxiality of the halo
could be attributed to the 
galaxy formation process itself (e.g., Dubinski \& Carlberg
1991; also see Binney 1996, and Rix 1996).  Thus the measurement of the disk
asymmetry allows one to constrain the details of galaxy
formation mechanisms. 

Since the disk and halo in a galaxy overlap and interact with each other 
gravitationally, the non-axisymmetry in one structural component will affect 
that in the other.   In the inner galaxy, inside of the
Holmberg radius, the disk constitutes a significant part of the
total mass of the galaxy, and the disk self-gravity can result in a substantial
decrease in the net non-axisymmetry of the halo potential as
shown for $m = 1$ by Jog (1999). Physically this is because the
disk self-gravity resists change and thus leads to a decrease in the magnitude 
of the net lopsided potential in the galactic plane. 
In this paper, we study a similar self-consistent, negative disk response to
higher order halo perturbations ($m =2 ,3$), and discuss the implications of 
our results for observations. A similar opposing disk response
for $m = 2$ has been mentioned   by Rix (1996),  and the decrease in 
ellipticity of the potential has been estimated approximately
by Binney (1996). 
Here we study this effect quantitatively, and
self-consistently, and find that the typical reduction in the
elliptical potential is  $\sim 20 \% $.

The orbits in a $m = 2$ and $m = 3$ perturbation potential are calculated
using the first-order epicyclic theory, and the density response of the
disk is obtained (\S 2).  Further, the self-gravitational potential
corresponding to the self-consistent density response of the disk to the halo 
potential is obtained by applying  to $ m = 2 $ and $3$  the general 
formalism developed by Jog (1999) (\S 2). The reduction factor 
for the halo potential due to the self-gravity of the disk is obtained for a 
wide range of galaxy parameters, including for the Milky Way, and the `true' 
halo asymmetry is obtained (\S 3). \S 4 contains a discussion of a few general 
points, and the conclusions from this paper are summarized in \S 5.

\section{POTENTIAL CORRESPONDING TO DISK 
     RESPONSE }

\subsection{Density Response In a Nonaxisymmetric Halo
  Potential}

We obtain the equations of motion for closed orbits, and the
density response of these, in an azimuthally symmetric galactic 
disk perturbed by a non-axisymmetric halo potential, 
and also obtain their relation to the isophotal shapes in an exponential
disk (see Appendix A). This is analogous to the lopsided case ($m=1$) 
studied by Jog (1997). We use the galactic cylindrical co-ordinates
$(R, \phi, z)$.

The unperturbed potential for the axisymmetric galactic disk, $\psi_0$, at a 
given radius $R$ is chosen to be a logarithmic potential which is
 applicable for a region of flat rotation, 
with $V_c$ being the constant rotational velocity:

$$ \psi_0 (R)  \: = \: {V_c}^2  ln R    \eqno (1) $$

The non-axisymmetric perturbation  halo potentials $\psi_{2}$ and $\psi_{3}$
corresponding to the $m=2$ and $3$ components are respectively
 chosen to be:

$$ \psi_{m} (R)  \: = 
    \: {V_c}^2 \: \epsilon_{m} \: cos m \: \phi   \: \:   \eqno (2) $$

\noindent where $\epsilon_{2}$ and $\epsilon_{3}$ are small, constant 
perturbation parameters, and $\phi$ is the  azimuthal angle in the plane of 
the disk. Note that the ellipticity, or the elongation, of the potential in 
the plane is given by $2  \epsilon_{2}$ and $2 \epsilon_{3}$ respectively for 
the $m=2$ and $3$ cases. The amplitude, ${V_c}^2 \: \epsilon_{m} $,
of the perturbation potential (eq.[2]) is assumed to be constant
with radius for simplicity, as in Franx \& de Zeeuw  (1992), and
 Rix \& Zaritsky (1995). This assumption is physically reasonable for a global 
 distortion of a halo as in a triaxial halo potential (see \S 1). We have also 
 assumed the phase of the perturbation potential to be constant with radius for
simplicity, as was done by Rix \& Zaritsky (1995) for $m =2$,
and by Jog (1997) for $m = 1$.  This is also justified for a global halo 
distortion assumed.

The unperturbed surface brightness of a typical galactic disk is
observed to have an exponential dependence on radius (Freeman 1970).
Assuming a constant mass-to-light ratio for the disk, this gives
the unperturbed mass surface density, $\mu_{un}$, of the stellar
disk to be:
 
$$\mu_{un} (R) \: = \: \mu_0 \: exp [- (R / R_{exp})]    \eqno (3) $$

\noindent where $\mu_0$ is the central extrapolated surface density, and
$R_{exp}$ is the scale length of the exponential disk. 

For the perturbed case due to the non-axisymmetric potential, the resulting 
isophote will be elongated along the 
same axis as the perturbed orbit, as argued by Franx \& de Zeeuw (1992). Since
the perturbed orbit, $\delta R$, is proportional to $cos m \phi$
(see e.g., eq. [A6] for $m=2$), the resulting effective surface 
density for the perturbed orbits in an exponential  disk  can be defined 
(see Rix \& Zaritsky 1995, Jog 1997) to be:

$$\mu (R, \phi ) \: = \: \mu_0 \: exp \left [ \: - \: \frac {R}{R_{exp}} \left
( 1 - \frac {(\epsilon_{iso})_m} {2} \: cos \: m  \phi \right ) \right ] $$

$$ \: \: \: \: \: \: \: \: \: \: = \mu_0 exp  [ \: - (R / R_{exp}) ] \:
    exp  [  (A_m / A_0) \: cos m \phi ] \:
 \eqno (4) $$

\noindent where  $(\epsilon_{iso})_m$  is the ellipticity of an
isophote at R for $m=2$ and $m=3$ respectively, and  $A_m / A_0$
is the fractional
amplitude of the $m$th   azimuthal Fourier component of the disk
surface brightness. Note that for small perturbations, this choice reproduces 
the definition of $(\epsilon_{iso})_m$ (eq.[A12]), as required.
The second relation on the r.h.s of  equation
(4) follows from the relation between $(\epsilon_{iso})_m$  and
$A_m / A_0$ (eq. [A13]) for an exponential disk.

Thus, the change in the  surface density, $\mu_{response}$, resulting 
from the
response of the disk to the perturbation halo potential is given by subtracting
$\mu_{un}$, the unperturbed surface density (eq. [3]), from eq.
(4), and for small perturbations  we obtain the density response for
$m=2,3$ to be respectively:

$$ [\mu_{response} (R, \phi)]_m \: = \: \mu_{un} (R) \:  
  \left (  \frac {A_m}{A_{0}} 
  \: cos m \: \phi  \right )    \eqno (5) $$

\noindent Thus, the disk response density is linearly proportional to
 $A_m / A_0$ or to the perturbation parameter, $\epsilon_m$ (see eq. [A15]), 
 as expected from the linear perturbation theory used in this paper.
Note that the response density is maximum along $\phi = 0^o $, 
along which the magnitude of the perturbation halo potential is also a maximum.
This result will be valid for any self-gravitating, centrally concentrated 
realistic disk mass distribution. 
  
\subsection{Disk Response Potential :  $ \vert m \vert
 $ = 2 , 3  Cases }

The self-gravitational potential $\psi$ (R, $\phi$, z)  for  
a general, nonaxisymmetric, thin disk with a surface density $\mu$ (R,
 $\phi$ , z) was obtained  by Jog (1999). This was obtained by solving the 
Poisson equation using the inversion technique involving the Hankel 
transforms of the potential-density pairs.  
The expression for the  nonaxisymmetric potential for
the thin disk is ( see eq. [22] from Jog 1999):

$$   \psi_{disk} \: (R, \phi, z) =  \: - G \:  \sum_{m = - \infty}^{\infty}
       \: exp \: (i m \phi) \: \: \int_0^{\infty}  \: 
        J_m (kR) \: exp (- k \vert z \vert) \:  dk 
	\int_0^{\infty} J_m (kR') \: R' \: dR' $$
       
$$     \qquad   \times \int_0^{2 \pi} \mu (R' , \phi ') \:
	\exp ( - i m \phi ') \: d\phi '    \eqno (6) $$

\bigskip

\noindent where $J_m(kR)$ is the cylindrical Bessel function of
the first kind, of order $m$. We apply this to the disk 
response density, $[\mu_{response} (R, \phi)]_m$ for $m=2$ and $m=3$ 
(eq. [5]). The resulting potential defines the response potential, 
$(\psi_{response})_m = [\psi_{response} (R, \phi)]_m$, respectively  
for $m = 2$ and $ m = 3$. We
consider the potential in the plane of the disk, so that $z = 0$.

Consider the case $\vert m \vert = 2$ first. Since the density response is 
proportional to $cos 2\: \phi$ ; only the values of $ m = \pm 2$ need to be 
kept in the integral over $\phi$ on the right hand side of equation (6),  
because only the contribution from these terms is non-zero. Further, since 
$cos 2\: \phi$ is
an even function of $\phi$, and since $J_{2} (kR) = - J_{-2} (kR)$;
hence the terms for $m=2$ and $m=-2$ contribute equally to the integral.
On substituting from equation (A15) for $A_2 / A_0$ in terms of ${\epsilon}_2$,
 the expression for the response potential, $(\psi_{response})_2$ ,
simplifies to:

$$ (\psi_{response})_2 \: = \: - \: 2 \pi G \: \mu_0 \: \epsilon_{2}
   \: cos 2\: \phi \: \: \int_{0}^{\infty} \:  J_2 (kR)   dk   $$

$$  \qquad \qquad \: \times \int_{0}^{\infty}
    \: J_2(kR') \: (1 + R' / R_{exp} ) \: \: exp (- \frac {R'}{R_{exp}}) R' dR'
                         \eqno (7) $$

The second integral over $R'$ can be solved using the relation (6.623.1)
from Gradshteyn \& Reyzhik (1980). The first integral over $R'$ can be
simplified by writing $J_2 (k R')$ in terms of 
 $J_1$ and $J_0$ using the standard recursion relation (e.g., Arfken 1970): 
$  J_0 (x) \: + \: J_2 (x) \: = \:  ({2}/{x}) \: J_1 (x)$ ; and the
 resulting two terms can be
solved using the relations (6.611.1) and (6.623.2) respectively from
Gradshteyn \& Reyzhik (1980). On substituting these in equation
(7), and writing $x = k R $, and setting $f(x) \equiv  x^2 \: (R_{exp} / R)^2
 \: + \: 1 $, equation (7)  simplifies to:

$$  (\psi_{response})_2 \: = \: - \: 2 \pi G \: \mu_0 
  \: R_{exp}  \: \epsilon_2 \: cos 2 \: \phi  \:  \: 
   (\frac {R}{R_{exp}})   \: \int_{0}^{\infty} \:   {J_2 (x)} dx $$

$$ \qquad \qquad \times   \left [ \frac {2}{x^2} \: \left (1 - \frac
{1}{[f(x)]^{1/2}} \right ) - \frac {(R_{exp}
/R)^2}{[f(x)]^{3/2}} \right ]   $$

$$ \qquad  \:   \: - \: 2 \pi G \: \mu_0 
  \: R_{exp}  \: \epsilon_2 \: cos 2 \: \phi  \:  \: 
    (\frac {R_{exp}}{R})^3  \: \: \int_{0}^{\infty} \: 
   \frac {J_2 (x)   3 x^2  \: dx} { [f(x)]^{5/2}}   \eqno (8) $$

\bigskip

\noindent  Note that $(\psi_{response})_2$ has {\it {a sign opposite to
the perturbation potential }} $\psi_{2}$ (eq. [2]), thus the
disk response is negative. In the linear
regime studied, the magnitude of the response potential is proportional to
$\epsilon_{2}$, the perturbation parameter in $\psi_{2}$,
and $cos 2 \: \phi $. Similarly, the disk response would be negative for any
self-gravitating, centrally concentrated disk. Following a similar analysis, 
the response potential $(\psi_{response})_3$,
is obtained for  $m = 3$, see Appendix B for details.

Next, define $\eta_m$ to be a dimensionless quantity as:

$$ (\eta_m) \: \: \equiv  \: \: \vert (\psi_{response})_m \vert \: /
  \: \psi_m       \eqno (9) $$

\noindent Note that this is independent of the strength of the imposed 
perturbation
potential, and depends linearly only on $\mu_0 \: R_{exp} \: / {V_c}^2 $.

The integrals over $x$ in equation (8) when solved analytically 
give terms involving hypergeometric series which need to be
calculated numerically. Instead,  equation (8) is directly solved numerically,
and while doing this   $J_2 (x)$ is obtained 
using a specific program (Press et al. 1986, Chap. 6) that gives a 
stable value for $n \geq 2$.

Define the dimensionless disk response potential, $\gamma_m$, as:
 
 $$ \gamma_m  \equiv  \: \eta_m   \:  \: \frac {{V_c}^2}
  {2 \pi G \mu_0 R_{exp} }   \eqno (10)  $$

In Figure 1, $\gamma_m $ vs. $R /R_{exp}$ is plotted for a flat
rotation curve  for $m=2$ and $3$. 
First, note that, at any radius, the magnitude of $\gamma_m$ is 
lower for the higher $m$ value. This follows
from the form of the response potential (eq.[6]), which
involves a double integral over the Bessel Function $J_m (x)$ which
decreases monotonically with $m$ for a particular argument $x$. 
Second, the maximum of $\gamma_m$ occurs at a lower radius with increasing $m$.
{\it {The maximum occurs}} at a radius =  1.42 $R_{exp}$ for $m=2$ and  
1.17 $R_{exp}$  for  $m=3$. For the corrected equations of motion for $m =1
$ (see Appendix A), the maximum occurs at 1.98 $R_{exp}$ which
is slightly larger than the radius of 1.4 $R_{exp}$ obtained in Jog (1999).
This radial dependence is a robust result valid for any exponential disk,
and is independent of the actual values of $\mu_0$ and $R_{exp}$.
The observational consequences of this result will be discussed
in \S $\: $ 4 .

\section{RESULTS}

\subsection{Net Nonaxisymmetric Potential: Self-Consistent
   Calculation}

A particle in the disk will be affected by the imposed halo
potential and also the disk response potential. We follow the
approach of Jog (1999) which gave a self-consistent calculation for $m
=1$. Thus, for a self-consistent case, the net nonaxisymmetric perturbation  
potential, $(\psi_{net})_m$,
in the disk plane is given by the sum of 
the  perturbation halo potential (eq. [2]), and the self-gravitational  
potential, $(\psi_{response})_m '$, which
corresponds to the disk response to the net potential,
$(\psi_{net})_m $. Thus,

$$ (\psi_{net})_m \: \equiv \: \psi_{m} \: + \: (\psi_{response})_m '
   \eqno (11) $$

\noindent In analogy with the disk response to the disk potential alone
(eq. [9]),
the net self-consistent response potential  $(\psi_{response})_m  ' $ is given 
to be:

$$   (\psi_{response})_m ' \: = \:  - {\eta}_m  (\psi_{net})_m
      \eqno (12)          $$

\noindent On substituting this in equation (11), we get 

$$   (\psi_{net})_m \: = \: \frac {\psi_m} { 1 + {\eta}_m }  \eqno (13) $$

\noindent Next, define the reduction factor, ${\delta}_m$, to be:

$$ \delta_m  \equiv \frac {1} { 1 + \eta_m }      \eqno (14) $$

\noindent Here $\delta (\leq 1)$  is the reduction or scaling factor by which 
the magnitude of $\psi_m$ is reduced due to the self-consistent, negative disk 
response. Note that since $\eta_m$ is a positive definite quantity, hence
$\delta_m \leq 1$. That is, the magnitude of the net
perturbation potential is always smaller than the magnitude of
the imposed halo perturbation potential. For $\delta_m = 1$, there
is no reduction and the disk response $\eta_m = 0$ as expected.
The reduction factor ${\delta}_m$ at a given radius $R/ R_{exp}$ 
 {\it is independent  of the strength of the perturbation potential } and 
 hence of $\epsilon_{m}$, and it   depends inversely on $\eta_m$ and hence 
inversely on $\mu_0 \: R_{exp} / {V_c}^2 $ . We will obtain 
the actual values of ${\delta}_m$ in \S 3.2 . 
$\delta_m$ will be a minimum at a radius where $\eta_m$ is a
maximum, that is at
1.42 and 1.17 $R/ R_{exp}$ for $m =2$ and $3$ respectively (see
Figure 1).

\subsubsection{Net Asymmetry}

Define the net, self-consistent,
perturbation potential, $(\psi_{net})_m$, in terms of  
a small perturbation parameter $(\epsilon_{net})_m$ to be:

$$(\psi_{net})_m \: \equiv \: {V_c}^2 \: (\epsilon_{net})_m \: cos \phi 
                      \eqno (15) $$
		      
\noindent Substituting this, and 
 $\psi_{m}$ (from eq. [2]) into equation (13), we obtain:

$$(\epsilon_{net})_m \: = \: \epsilon_{m} \:  (\delta)_m   \eqno (16) $$

\noindent Thus, the parameter
$\epsilon_{net}$ denoting the strength of the net perturbation potential in the
galactic disk is reduced compared to $\epsilon_{m}$, the parameter
denoting the perturbation halo potential, by the reduction factor 
$\delta_m$ .
Observations of $A_m / A_0$, the fractional amplitude of the $ m^{th}
$ azimuthal Fourier component of the surface brightness, will yield  
the parameter $(\epsilon_{net})_m$. Here $ 2(\epsilon_{net})_m$ is
the net ellipticity of the halo potential.  Hence the 
halo-alone case (eq. [A15]) is now modified, and we get the net
ellipticity to be:

$$( \epsilon_{net})_2  \: = \: \frac {A_2 /A_0} { (1   
          + \frac {R}{R_{exp}}  ) }      \eqno (17) $$

 Similarly, 
$({\epsilon_{net}})_3 $ is given by r.h.s. of equation (A7).
 The values of the true ellipticity 
($ 2 \epsilon_2$) will be obtained in \S 3.3.
 
\subsection{Reduction Factor, ${\delta}_m $}

The value of the reduction factor ${\delta}_m $ (eq.[14]) is obtained 
numerically  and its variation with the galaxy morphological type, size, and 
radius in the galactic disk, and the component $m$, is studied for the  
classical large or giant spiral galaxies.

\subsubsection{Giant Spiral Galaxies}

The values of the typical disk parameters for the giant spiral galaxies
 are taken to be: $\mu_0$, the central surface mass
density $  = 450 M_{\odot} pc^{-2} $;
$R_{exp}$, the exponential disk scale length $ = 3 kpc$; and a range of 
values of the maximum  rotation velocity, $V_c$, for a flat
rotation curve are taken to be  
equal to 200, 250, and 300  $km s^{-1}$. See Jog (1999) for a discussion
supporting the choice of these values.
Figure 2 contains a plot of the reduction factor, $(\delta)_2$, 
 by which the elliptical halo potential is reduced due to
the negative response of the disk (eq. [14]), vs. $R / R_{exp}$. Note that 
 the reduction factor is a minimum
at $1.42 R_{exp}$ and increases thereafter, as expected from equation (14)
(see \S 3.1). The typical minimum value of  $\delta$
 lies in the range of 0.75 - 0.9 , and $\delta $ is larger for galaxies 
 with a larger value of $V_c$ .

A similar plot of $\delta_3$ versus radius (not shown here)
gives the minimum value of $\delta_3$ to be higher, in the range of
$0.83 - 0.93 $.
Thus the reduction due to the disk self-gravity in the perturbation
halo potential is not important for $m = 3 $, and for the higher
values of the harmonic $m$. Therefore, for $m \geq 3$, the observed 
asymmetry in the disk response represents the true halo asymmetry.

For the corrected equations of motion for the lopsided case ($ m
=1$, see Appendix A), the resulting minimum in $\delta$ is found
to span a
range of 0.67 - 0.82 and it occurs at a radius of $ 1.98 R_{exp}$.
These are slightly different from the results of Jog (1999), mainly
in the peak radius which was earlier obtained to be $1.4 R_{exp}$. Note,
however, that the correct results obtained here are still in good
agreement with the observed radial variation in the net lopsided 
distribution, which was a main result of Jog (1999).

\subsubsection{The Milky Way}
 
Next consider the special 
case of a giant spiral galaxay, namely,  the Milky Way Galaxy. We assume a 
flat rotation curve with $V_c = 220 km s^{-1} $, $R_{exp} = 3.5 kpc $ 
(Binney \& Tremaine 1987), and
$\mu_0 = 450 M_{\odot} pc^{-2} $ as discussed above for the giant
galaxies. The plots of   ${\delta}_2$ and ${\delta}_3$
vs. $R / R_{exp}$ are given in Figure 3.
 Several quantitative resuls follow from this. First, the
minimum values of ${\delta}_2$ and ${\delta}_3$ are 0.75 and
0.88 respectively. Second, ${\delta}_2$ is $0.79$ in the solar neighborhood of 
$R =8.5 kpc$, which is $= 2.43 \times R_{exp}$. This  is almost twice the
 value of $3/7$ for the reduction in the ellipticity
 estimated from an order-of-magnitude calculation by Binney (1996). 
Because of the general formulation in our paper, Figure 2 gives 
 the reduction factor as a function of radius for $m=2$ for a variety of 
 galaxy parameters.

Third, if the halo of our Galaxy has an elliptical halo potential, 
then the disk response would reduce this potential at most by a factor of 
$\sim 0.75 $. Thus for the observed values of the disk parameters for the 
Galaxy and also the other galaxies, while the negative disk response cannot be
ignored, it  can never totally cancel or counteract the imposed elliptical 
halo potential in the disk plane. This is contrary to the suggestion by 
Binney (1996) that at high enough disk to halo mass ratio, the galaxy could be 
treated as axisymmetric.

\subsection{Ellipticity of the Halo Potential }
 
The true ellipticity of the halo potential is an important physical
property, possibly related to the process of galaxy formation
(\S 1), and
attempts have been made in the literature to estimate this from
the observational data on disks of galaxies. The
resulting values span a large range. The optical data
on the elongation in the disk yield typical estimate of ellipticity for the 
halo of spirals to be 0.1 (Franx \& de Zeeuw 1992), while a
value of 0.045 is obtained by a
similar analysis of the near-infrared study of a smaller sample
of 18 galaxies (Rix \& Zaritsky 1995). From detailed kinematical studies,
an ellipticity of 0.1 is obtained for the Milky Way Galaxy (Kuijken \& 
Tremaine 1994), and for NGC 2403 and NGC 3198 this is estimated respectively 
to be 0.064 and 0.019 (Schoenmakers et al. 1997). From the scatter in the 
Tully-Fisher relation, the halo ellipticity is estimated to be 0.1 (Franx 
\& de Zeeuw 1992).

The observations measure the net ellipticity,  $2 (\epsilon_{net})_2$, while 
the $\it true$ ellipticity of the halo potential,
 given by $2 \epsilon_2 $,  is higher by a factor of $ 1 / {\delta}_2 $
(eq.[16]). Since the typical minimum value of $\delta_2$ is 0.8
(Figure 2),  the true ellipticity is higher by 20 \% than the measured 
net value. Thus, for the above observed typical range of net ellipticity
of $0.045 - 0.1 $ in the literature, the true halo ellipticity is
in the range of 0.056 - 0.12 .  This is an important new physical result 
from our work. Of course, the above estimates would have
substantial error bars due to the contamination by the spiral
arms or a central bar. 

Note that the reduction factor $\delta_2 \rightarrow 1 $ at large $R
\geq 8 R_{exp}$ (Figs. 2-3), or about twice the Holmberg radius. Hence
 the true halo ellipticity can be directly sampled by 
 studying the tracer at larger radii namely atomic hydrogen gas.
This was done in the plane of IC 2006 (Franx, Van Gorkom, \& de
Zeeuw 1994), and the halo was found to be axisymmetric. The
ellipticity perpendicular to the plane of the galactic disk could be obtained
by studying the polar ring galaxies as suggested by  Rix (1996).

\section{DISCUSSION}

1. The net nonaxisymmetry in the disk will only manifest beyond the radius 
where the magnitude of the disk repsonse potential is a maximum,
and its  magnitude will increase with radius as seen from the definition of
 $\delta_m$ (\S 3.1), and this radius is larger for a lower $m$.
  This indicates the increasing relative dynamical 
importance of halo over the disk at large radii.
   Therefore, if the halo distortion is constant or increasing with
radius, then the disk lopsidedness ($m =1 $) would be seen only in the outer
disk while the higher-order nonaxisymmetric features
could be seen farther in the disk.
However, the radial variation for $m \ge 2 $
in the inner/optical region will be affected by spiral arms, and bars. 
 Hence the radial variation in $m \geq 2$ components cannot be given clearly, 
 in contrast to the $m=1$ case where a clear minimum radial distance was 
 predicted for the detection of $m=1$ global features (Jog 1999,
also \S 3.2).

2.  In addition to the disk response to the global perturbation in the halo 
 as studied in this paper, there could also be $m=2$ or $m=3$ modes 
generated directly in the disk, say due to gravitational instabilitites. These
would typically have a strong phase variation with radius and hence be 
detected as the standard two-armed or three-armed spiral features 
respectively. Only future detailed simulations of tidal
interactions on lines of Wienberg (1995) will tell
us about the strength as well as the phase and the
radial dependence of the true  halo non-axisymmetry of the
various $m$ components. Since there is no Inner Lindblad
Resonance for $m =1$ in a typical galactic a disk (e.g., Block et al. 1994),
the $m=1$ component may dominate in the non-linear regime.

3. The negative disk response decreases the net asymmetry of the potential 
in the galactic plane, and this would affect the further evolution of the 
galaxy. This could be one reason why numerical simulations with a higher
mass concentration in the form of a disk show a decrease in the halo
ellipticity (Dubinski 1994). Thus the
negative feedback due to the self-gravity of the disk highlighted in
this paper should be included in future studies of galaxy evolution.
The present paper shows that  the disk cannot be treated as a collection of
massless test particles - doing so would overestimate the disk
reponse.

\bigskip

\section{CONCLUSIONS}

We have calculated the self-consistent disk response for an axisymmetric 
galactic disk perturbed by a non-axisymmetric halo potential with elliptical 
and higher-harmonic perturbations ($m =2$ and $3$) :

1. The self-gravitational potential of the
self-consistent density response of a galactic disk
is calculated and this is shown to oppose the perturbation potential. 
Thus,  the magnitude of the net non-axisymmetric potential in
the galactic disk plane is always reduced compared to that of
the perturbation potential. This reduction is denoted by a
factor, $\delta_m$, which is found to be independent of the strength of the 
perturbation potential.

2. The reduction factor, $\delta_2 $, is obtained for a wide range of galaxy 
parameters, including  for the Milky Way. It has a minimum
value of $\sim  0.75 - 0.9 $,  
which is insensitive to the morphological type and size of the galaxy.
The reduction is most significant at 1.4 disk scale lengths and is
less important at higher radii.
Beyond twice the Holmberg radius,
the reduction is negligible and the atomic hydrogen gas can be
used to trace the true ellipticity of the halo potential.
 In the solar neighborhood of the Milky Way,  the elliptical halo 
potential is decreased by a factor of $\delta_2 = 0.8 $ due to the disk 
self-gravity. The reduction is negligible for the higher
harmonics ($m \geq 3 $) of the halo potential. The asymmetric
disk response in $m \geq 3$ therefore represents the true halo asymmetry.

3.  On correcting for the negative disk response, the {\it true} ellipticity
of the halo potential for a typical spiral galaxy is shown to be
higher by $\sim 20 \% $ than the halo ellipticity of $\sim 0.5 -
0.1$ deduced in the literature from observations of isophotal or 
kinematical asymmetry of disks.

4. The negative disk response due to the disk self-gravity is
shown to be always significant in decreasing the strength of the
non-axisymmetric halo potential. Hence the galactic disk in a realistic
galaxy cannot be treated as a collection of massless test particles.
Yet, the negative disk response in a real galactic disk is found to be never 
large enough to completely obliterate the effect of the halo ellipticity.

\bigskip

\bigskip

I would like to thank the anonymous referee for comments that
led to a clearer presentation of material in Section 2, and the
Appendix A.

\newpage

\appendix

\section{\bf APPENDIX A - PERTURBED ORBITS AND ISOPHOTAL SHAPES }

We study the orbits in an axisymmetric disk perturbed by a halo
potential of harmonic numbers $m =  2 $, and $3$. We use
the cylindrical co-ordinate system ($R, \phi$),
where $\phi$ is the azimuthal angle in the galactic plane.
Consider a perturbed orbit around the initial circular orbit
at a radius $R_0$, which is given by $R = R_0 + \delta R$ and
$\phi = \phi_0 + \delta \phi$. Here $\phi_0 = \Omega_0 t$, where
$\Omega_0$ is the circular rotation speed at $R_0$ and is given by
the following where $\psi_0$ is the unperturbed potential:

$$ R_0 {\Omega_0}^2 \: = \: \frac {d \psi_0}{d R} \vert_{R_0}
              \eqno (A1) $$

Consider the $m = 2$ case first. The general perturbation potential
is taken to be $\psi_{pert} (R_0) \: cos 2 \phi_0 $.
Following the procedure for the first-order epicyclic theory as in
Jog (1997), the general coupled equations of motion for $\delta R$ and
$\delta \phi$ are given by equations (4) and (5) from Jog
(1997) to be:

$$ \frac {d^2 \delta R} {d t^2} = \: - \: \delta R \: \: \left(
3 {\Omega_0}^2 \: + \: \frac {{d}^2 \psi_0} { d
R^2}   \: \: \vert_{R_0}  \: \right ) \: \: - \: \: 
\left ( \frac {2 \psi_{pert} (R_0)} {{R_0}} \: + \: 
 \frac {d  \psi_{pert}} {d R} \: \: \vert_{R_0}  \: \right)
 \: cos 2 \phi_0            \eqno (A2) $$

$$ R_0 \frac {d^2  \delta \phi} {d t^2} \: + \: 2 \Omega_0  \: 
   \frac { d  \delta R} {d t} \:
   = \:  \frac {2 \psi_{pert} (R_0)} {R_0} \: sin 2 \phi_0     \eqno (A3) $$

From the theory of a forced oscillator (e.g., Symon 1960), equation (A2) may
 be solved to yield the following solution for the closed orbits:

$$ \delta R = \frac {- \left [ \:  \frac{2 \psi_{pert} (R_0)} {R_0}
  \: + \: \frac {d  \psi_{pert} } { d R} \: \:  \vert_{R_0}  \: 
  \right] }     
  { {\kappa}^2 - 4 {\Omega_0}^2 } \: \: \: cos 2 \phi_0
                               \eqno (A4) $$

\noindent where $\kappa$ is the epicyclic frequency at $R_0$.

 For the present problem, $\psi_{pert} \: = \: {V_c}^2 \: 
\epsilon_2 \: $ is the amplitude of the perturbation potential $\psi_2$ defined
by equation (2), and the unperturbed disk potential $\psi_0$ is
defined by equation (1). For these,  equation (A4) gives:

$$ \delta R =  R_0  \:  \epsilon_{2} \: cos 2 \phi_0  \eqno (A5) $$

\noindent Thus the net radius is given as:

$$ R \:  =  \: R_0  \: +  \: \delta R  \: = \: R_0 \: ( 1 \: + \:  
      \epsilon_{2} \: cos \: 2 \phi_0)  \eqno (A6) $$

\noindent Hence $V_R$, the perturbed velocity along the radial direction,
is given as:

$$V_R \: =  - \: 2 \: V_c \: \epsilon_{2} \: sin 2 \phi_0  \eqno (A7) $$

On substituting the solution for $\delta R$ from equation (A5) in
equation (A3) for the $\phi$ component of the equation of motion
and integrating it we obtain the solution for the perturbed azimuthal
velocity component $R_0 \:  d (\delta \phi) / dt $ . 
The net velocity  along the azimuthal direction 
is obtained to be:

$$  V_{\phi}  \: = \: V_c \: + \: R_0 \:  d (\delta \phi) / dt \: + \:
\Omega_0 \: \delta R  = V_c \: ( 1 -
   2 \: \epsilon_{2} \: cos \: 2 \phi_0 )           \eqno (A8)  $$

Hence the equations of motion for the perturbed, closed orbits
in the $m=2$ perturbed potential are given by equations (A6)-(A8).

A similar procedure for $m = 3$ yields the following equations of motion 
for the perturbed, closed orbits:

$$ R = R_0 [ 1 + (2/7)  \epsilon_{3} cos 3 \phi_0 ] \: \: ; \: \:
 V_R = - (6/7) V_c {\epsilon_{3}} sin 3 \phi_0 \: \: ; \: \:
 V_{\phi} = V_c  [ 1 - (9/7) \epsilon_{3} cos 3 \phi_0 ]  \eqno (A9) $$

The results for the $m=1$ (lopsided) case from Jog (1997) are given below 
for comparison. The change in the azimuthal velocity, 
$\delta V_{\phi} $, was given in Jog (1997) to be 
equal to $R_0 \: d (\delta \phi) / dt $. The correct, space-frame velocity 
(see Schoenmakers 1999), would include an additional term = $
\Omega_0 \: \delta R $ (see for example the l.h.s. of eq. [A8]). 
On including this correction, the perturbation term in 
$V_{\phi}$ is now changed and has a  factor
1 instead of 3 in it,  while the expressions for $R$ 
and $V_R$ remain the same as in Jog (1997). The revised equations
of motion for the $m =1$ case are given here for the sake of
completeness, and are:

$$ R = R_0 ( 1 - 2  \epsilon_{1} cos  \phi_0 ) \: \: ; \: \:
 V_R = 2 V_c {\epsilon_{1}} sin  \phi_0 \: \: ; \: \:
 V_{\phi} = V_c  ( 1 + \epsilon_{1} cos  \phi_0 )  \eqno (A10) $$

Next, we study  the resulting isophotal shapes for an exponential disk. 
This analysis is similar to the $m
= 1$ case studied by Jog (1997). For an exponential galactic
disk (eq.[3]), $A_m / A_0 $, the fractional amplitude of the $m^{th}$
 azimuthal Fourier component of the surface brightness is obtained to be:

$$A_m/A_0 \equiv \Delta \mu / <\mu>
 =  \vert \: - \:   \frac {\Delta R} {R} 
   \: \frac {R} {R_{exp}} \vert   \eqno (A11) $$

\noindent where $<\mu>$  is the azimuthal average of the disk density.
 Here $ \Delta R  / R $ is the distortion in the isophote and is related to 
 $(\epsilon_{iso})_m$, the ellipticity of an isophote at $R$, as follows:

$$ (\epsilon_{iso})_m \: \equiv 1 - ( R_{min} / R_{max} ) = \: 2  \: 
    ( \Delta R / R)          \eqno (A12)  $$

\noindent where $R_{min}$ and $R_{max}$ are the minimum and
maximum extents of an isophote respectively.

From equations (A11) and (A12), we get:

$$ A_m/A_0 =  \frac {(\epsilon_{iso})_m} { 2}  \: \frac {R} {R_{exp}}
           \eqno (A13) $$

We next obtain a relation between the
perturbation parameter, $\epsilon_m$, and the resulting  $A_m / A_0  $. 
Since the orbital velocity changes along the perturbed orbit, the
associated surface density also changes as a function of the
angle $\phi$ . The changes for particles on these orbits are governed  by the
equation of continuity which has the following form in
cylindrical co-ordinates:

$$\frac {\partial} {\partial R} \left [ R \: \mu (R, \phi) \:
 V_R (\phi) \right ] \: \: + \: \: \frac {\partial} {\partial \phi}
  \left [ \mu (R, \phi) \:  V_{\phi} (\phi) \right ] \: = \: 0
     \eqno (A14) $$

On solving together the equations of the perturbed motion as
given by equations
[A6]-[A8] for $m=2$, or equation [A9] for $m = 3$, or equation [A10] for 
$ m = 1$; and the continuity equation (eq. [A14]); and the equation  for
the effective surface density (eq. [4]); we obtain the  relation 
between $(\epsilon_{iso})_m$, the ellipticity of an isophote,  and 
$\epsilon_{m}$, the perturbation parameter for the potential, at a given radius
$R$. On combining this with equation (A13), we get the relation
between $\epsilon_m$ and $A_m / A_0 $, the fractional amplitude
of the azimuthal Fourier component, for the $m = $ 1, 2, and 3 cases
to be respectively:

$$   \epsilon_{1}  \: = \: \frac {A_1 /A_0} { (   
          2 R/  R_{exp}  ) - 1 }  \: \: ; \: \: \:
     \epsilon_{2}  \: = \: \frac {A_2 /A_0} { ( 1  
          + R/  R_{exp}  ) } \: \: ; \:  \: \: 
	  \epsilon_{3}  \: = \: \frac {A_3 /A_0}  { 1 +  
           \frac {2  R} { 7 R_{exp} } }      \eqno (A15) $$

This is used to write the disk density response (eq.[5])
and hence the response potential (\S 2.2) in terms of $\epsilon_m$,
the perturbation parameter of the potential.

\newpage

\appendix

\section{APPENDIX B - DISK RESPONSE POTENTIAL : $\: \vert m \vert = 3 $ 
   CASE}

We obtain the disk response potential, $(\psi_{response})_3$
defined in \S 2.2 , starting from eq. [6].
Following the reasoning  as in \S 2.2 ,
only the terms for $m \pm 3$ need to be kept, and using the
expression for $A_3 /A_0$ (eq.[ A15]), we obtain :

$$ (\psi_{response})_3 \: = \: - \: 2 \pi G \: \mu_0 \: 
   \epsilon_3  \: cos 3\: \phi
 \: \: \int_{0}^{\infty} \:  J_3 (kR)   dk   $$
 
$$  \qquad \qquad  \times  \: \int_{0}^{\infty}
    \: J_3(kR') \: [ 1 + (2/7) (R' / R_{exp} ) ] \:
     exp (- \frac {R'}{R_{exp}}) R' dR'                     \eqno (B1) $$

Next, we use the
recursion relations between $J_3$ and the lower order Bessel functions,
$J_0$ and $J_1$ (e.g., Arfken 1970) to simplify the integrals over $R'$, 
and solve the various terms by applications of the relations
(6.623.1, .2, and .3)
and (6.611.1) from  Gradshteyn \& Ryzhik (1980). Next write $x =
kR$,  and define the function  $f(x) \equiv {x^2 (R_{exp} / R)^2 \: + \: 1 }$;
and obtain:

$$    (\psi_{response})_3    \: = \: - \: 
   { 2 \pi G \mu_0 R_{exp}} \:  \epsilon_3 \: cos 3 \phi  \: \: 
     (R_{exp} / R)^2 \:
      \int_{0}^{\infty} \: {J_3} (x) \:   dx  $$
      
$$ \qquad  \times       \: \left ( \frac {- x}
     {(f(x))^{3/2}} \: \: + \: \: \frac
{8 ([f(x)]^{1/2} - 1 )} 
    {x^3 \: (R_{exp} /R)^4 }   \: \: 
 - \: \: \frac{4 (R/ R_{exp})^2} {x \: [f(x)]^{1/2}} \right )  $$
 
$$ \qquad \qquad  \qquad -   { 2 \pi G
  \mu_0 R_{exp} } \: \: \epsilon_3 \: \: cos 3 \phi \: \: \frac {2}{7} \: 
  \:   \int_{0}^{\infty} \: {J_3} (x) \:   dx $$

$$\times  \left [ - \frac {3x \: (R_{exp} /R)^2} {(f(x))^{5/2}} \:   + \:
  \frac { 8 (R / R_{exp})^2}{x^3}
       \left( 1 - \frac {1}{(f(x))^{1/2}} \right) 
     - 
     \frac {4}{ x \: (f(x))^{3/2} }  \right ]
      \eqno (B2) $$

\newpage

\centerline{\bf REFERENCES}

\bigskip

\noindent Arfken, G. 1970, Mathematical Methods for Physicists
(2d ed.; New York : Academic)

\noindent Baldwin, J.E., Lynden-Bell, D., \& Sancisi, R. 1980, MNRAS,
           193, 313

\noindent Binney, J. 1978, MNRAS, 183, 779

\noindent Binney, J.  1996, in  Unsolved Problems Of The
 Milky Way, IAU Symposium 169, ed. L. Blitz \& P. Teuben
 (Dordrecht: Kluwer), 1

\noindent Binney, J.,  \& Tremaine, S. 1987, Galactic Dynamics,
           (Princeton : Princeton University Press)

\noindent Block, D.L., Bertin, G., Stockton, A., Grosbol, P.,
  Moorwood, A.F.M.,  \& Peletier, R.F. 1994, A \& A, 288, 365

\noindent Dubinski, J. 1994, ApJ, 431, 617

\noindent Dubinski, J., \& Carlberg, R.G. 1991, 378, 496

\noindent Franx, M., \& de Zeeuw, T. 1992, ApJ, 392, L47
 
\noindent Franx, M., van Gorkon, J.H.,  \& de Zeeuw, T. 1994, ApJ, 436, 642 
 
\noindent Freeman, K.C. 1970, ApJ, 160, 811

\noindent Gradshteyn, I.S., \& Ryzhik, I.M. 1980, Tables of Integrals,
 Series, and Products (New York: Academic Press)

\noindent Haynes, M.P., Hogg, D.E., Maddalena, R.J., Roberts,
 M.S., \& van Zee, L.  1998, AJ, 115, 62
 
\noindent Jog, C.J. 1997, ApJ, 488, 642

\noindent Jog, C.J. 1999, ApJ, 522, 661

\noindent Kornreich, D.A., Haynes, M.P., \& Lovelace, R.V.E.
1998, AJ, 116, 2154

\noindent Kuijken, K., \& Tremaine, S. 1994, ApJ, 421, 178

\noindent Press, W.H., Flannery, B.P., Teukolsky, S.A., \&
  Vetterling, W.T. 1986, Numerical Recipes (Cambridge: Cambridge
  Univ. Press), chap. 6.

\noindent Richter O.-G., \& Sancisi, R. 1994, A \& A, 290, L9

\noindent Rix, H.-W.  1996, in  Unsolved Problems Of The
 Milky Way, IAU Symposium 169, eds. L. Blitz \& P. Teuben
 (Dordrecht: Kluwer), 23

\noindent Rix, H.-W., \& Zaritsky, D. 1995, ApJ, 447, 82 $\: \:$

\noindent Rudnick, G., \& Rix, H.-W. 1998, AJ, 116, 1163

\noindent Sandage, A. 1961, The Hubble Atlas Of Galaxies
 (Washington: Carnegie Institution Of Washington)

\noindent Schoenmakers, R.H.M. 1999, Ph.D. thesis, University of
 Groningen.

\noindent Schoenmakers, R.H.M., Franx, M.  \& de Zeeuw, P.T. 1997, MNRAS, 
  292, 349

\noindent Symon, K.R.  1960,  Mechanics (Reading: Addison-Wesley)

\noindent Weinberg, M.D. 1995, ApJ, 455, L31

\noindent   Zaritsky, D. \&  Rix, H.-W.  1997, ApJ, 477, 118

\newpage

\centerline {\bf FIGURE CAPTIONS}

\figcaption[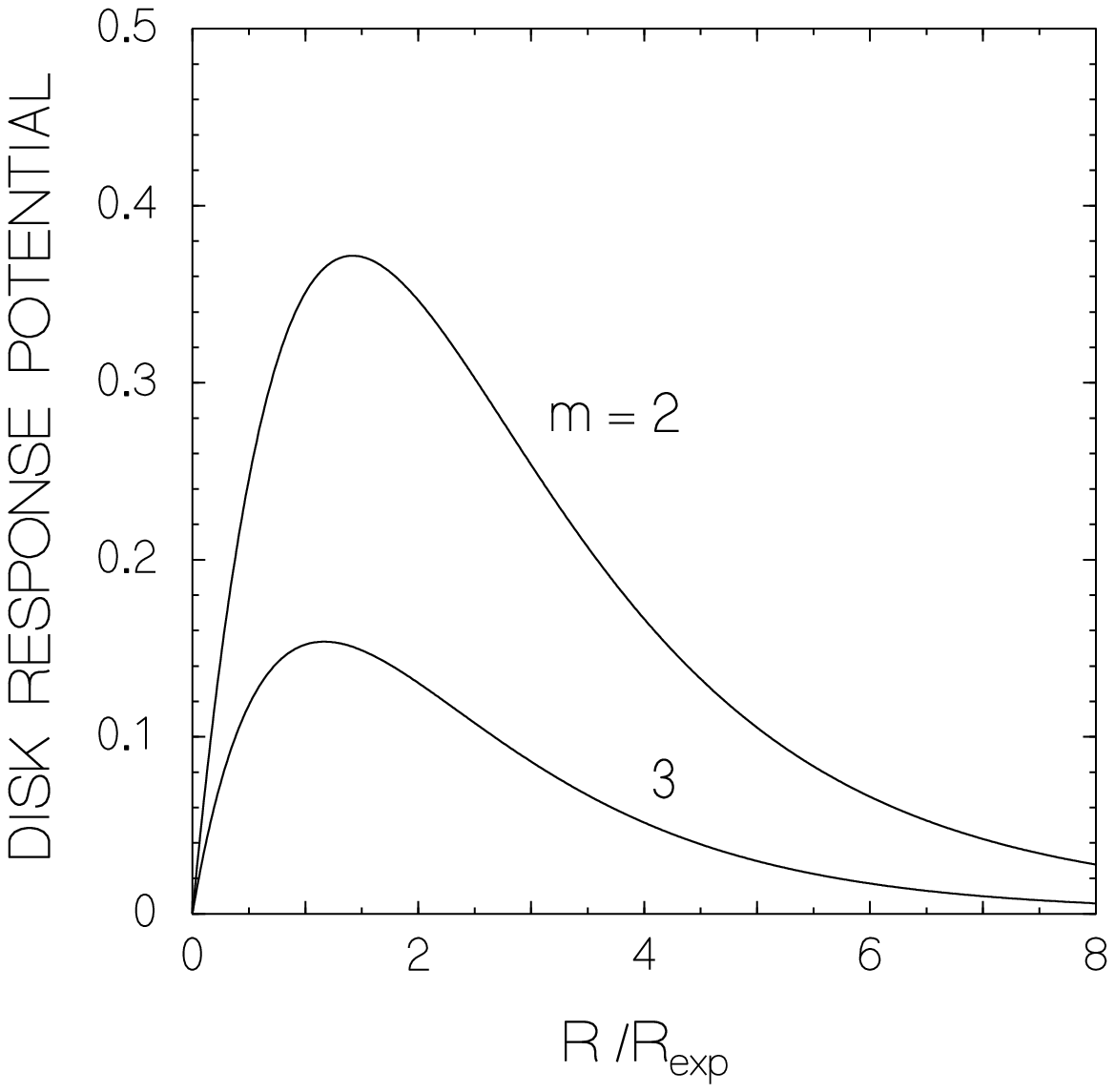]{ The dimensionless 
self-gravitational potential of the
disk response, $(\gamma)_m$
 versus  the dimensionless radius $ R / R_{exp}$  for  
the azimuthal wavenumber, $m$ = 2 and 3.
The  maximum occurs at a radius of  1.42 and 1.17 exponential
disk radii respectively for $m =  2$ and $ 3 $.\label{fig1}}

\figcaption[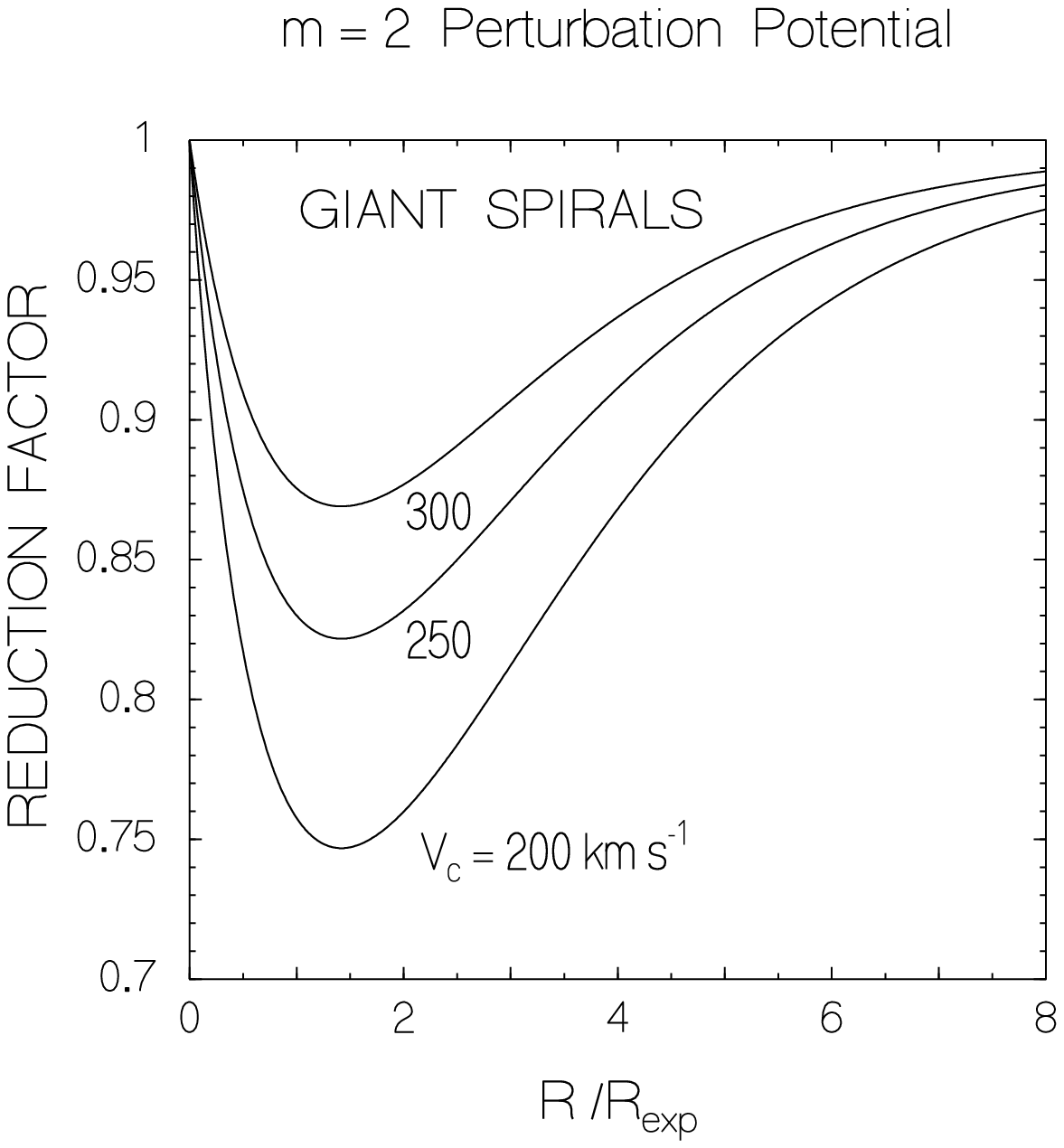] { The reduction factor $\delta$ due to the
 self-consistent, negative disk response for the $m=2$ perturbation halo 
 potential versus the radius $R / R_{exp}$, for giant spiral galaxies, with a 
 flat rotation curve with a velocity $V_c = 200, 250, $ and $300 km s^{-1}$,
and $R_{exp}$ = 3 kpc. The minimum reduction factor lies in the
range of 0.75 - 0.9 , and 
it always occurs at $R / R_{exp}$ = 1.42, and $\delta$ increases
steadily beyond this radius.\label{fig2}}

\figcaption[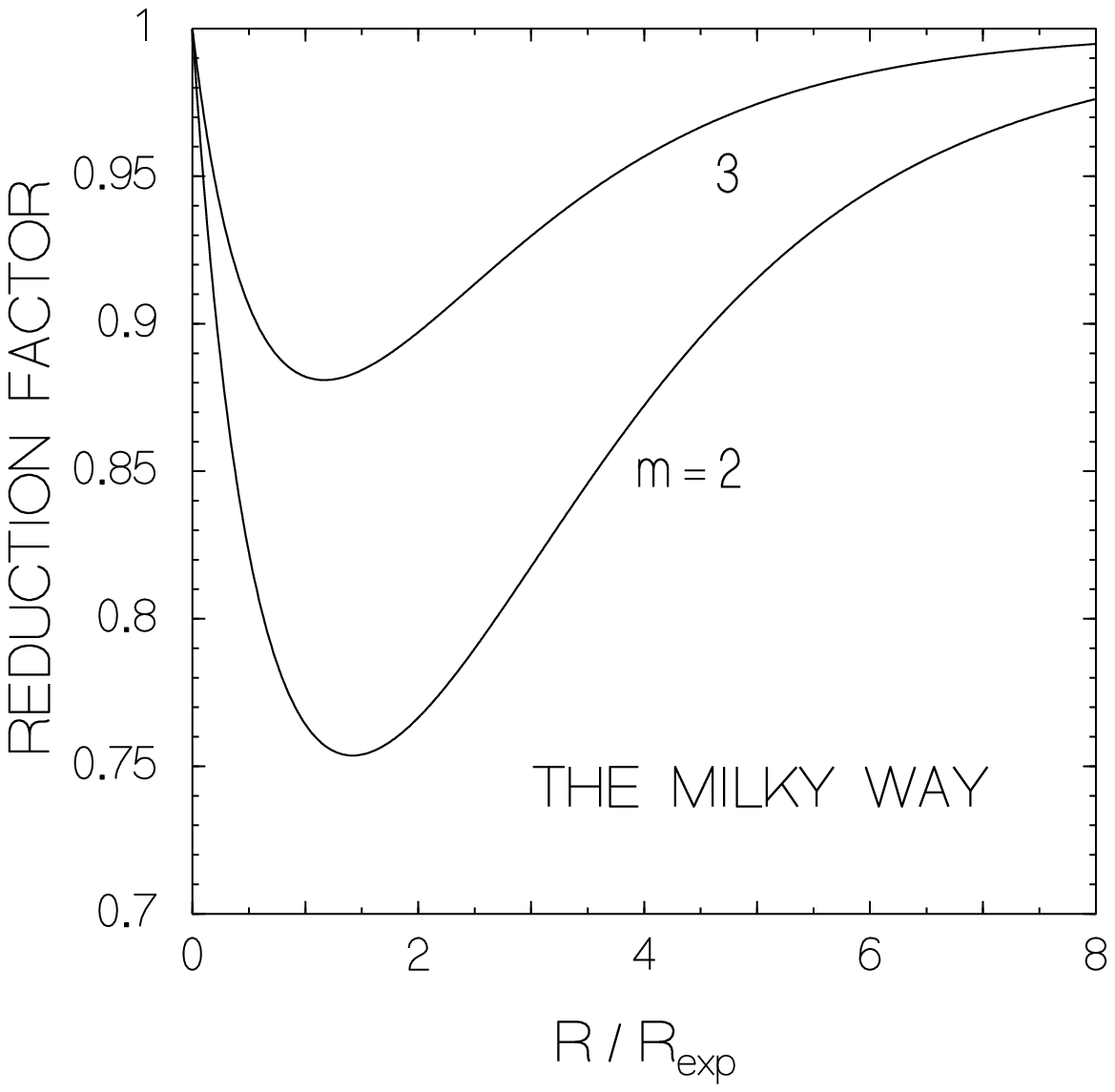] { The reduction factor $\delta$
versus the radius $R / R_{exp}$, for the Milky Way, with a flat
rotation curve with a velocity $V_c = 220 km s^{-1}$, and
$R_{exp}$ = 3.5 kpc, for $m = 2$ and $3$.
The minimum reduction factor is $0.75$ and $0.88$, similar to 
typical giant spiral galaxies (see Fig. 2); and occurs at 1.42
and 1.17 $R_{exp}$ for $m$ = 2 and 3 respectively.\label{fig3}}

\end{document}